\documentclass[sigconf]{acmart}

\pdfoutput=1

\usepackage[utf8]{inputenc}
\usepackage[most]{tcolorbox}

\usepackage{booktabs}

\usepackage{graphicx}
\usepackage{multirow}
\usepackage{setspace}
\usepackage{amsmath}
\usepackage{amsfonts} %
\usepackage{calligra} %
\usepackage{mathtools}

\usepackage{amsthm} 

\usepackage{comment}
\usepackage{listings}
\usepackage{enumitem}

\usetikzlibrary{patterns}

\usepackage{blindtext}

\usepackage{acronym}

\usepackage{float}
\usepackage{textcomp}

\usepackage{algorithm}
\usepackage{algorithmic}

\usepackage{url}

\usepackage[all=normal,paragraphs=tight, floats=tight,indent=tight]{savetrees}
\usepackage{framed}
\usepackage{soul}
\usepackage{csquotes}
\usepackage{multirow}
\usepackage{array}
\newcolumntype{L}[1]{>{\raggedright\let\newline\\\arraybackslash\hspace{0pt}}m{#1}}
\newcolumntype{C}[1]{>{\centering\let\newline\\\arraybackslash\hspace{0pt}}m{#1}}
\newcolumntype{R}[1]{>{\raggedleft\let\newline\\\arraybackslash\hspace{0pt}}m{#1}}
\newcolumntype{H}{>{\collectcell\lstinline}l<{\endcollectcell}}
\usepackage{url}
\MakeOuterQuote{"}
\usepackage[caption=false,font=footnotesize,labelformat=simple]{subfig} 

\newcommand{\sol}{DAVE}

\definecolor{mygreen}{rgb}{0,0.6,0}
\definecolor{mygray}{rgb}{0.5,0.5,0.5}
\definecolor{mymauve}{rgb}{0.58,0,0.82}

\acrodef{CPS}{Cyber-Physical System}
\acrodef{IoT}{Internet of Things}
\acrodef{HDL}{Hardware Description Language}
\acrodef{CAD}{Computer-Aided Design}
\acrodef{EDA}{Electronic Design Automation}
\acrodef{HPC}{High-Performance Computing}
\acrodef{DL}{Deep Learning}
\acrodef{ML}{Machine Learning}
\acrodef{NLP}{Natural Language Processing}
\acrodef{IC}{Integrated Circuit}
\newcommand{\ignore}[1]{{}}

\newcommand{\squishlist}{
	\begin{list}{$\bullet$}
		{ \setlength{\itemsep}{0pt}
			\setlength{\parsep}{1pt}
			\setlength{\topsep}{1pt}
			\setlength{\partopsep}{0pt}
			\setlength{\leftmargin}{0.9em}
			\setlength{\labelwidth}{1.5em}
			\setlength{\labelsep}{0.4em} } }
	\newcommand{\squishend}{
	\end{list}  }

\definecolor{graphFirst}{RGB}{2,136,209} 
\definecolor{graphSecond}{RGB}{211,47,47} 
\definecolor{graphThird}{RGB}{245,124,0} 
\definecolor{graphFourth}{RGB}{56,142,60} 
\definecolor{graphFifth}{RGB}{81,45,168} 
\definecolor{graphSixth}{RGB}{69,90,100} 
\definecolor{graphSeventh}{RGB}{251,192,45} 
\definecolor{backgroundSecond}{RGB}{239,154,154} 
\definecolor{backgroundThird}{RGB}{255,204,128} 
\definecolor{backgroundFourth}{RGB}{165,214,167} 
\definecolor{backgroundFifth}{RGB}{179,157,219} 
\definecolor{backgroundSixth}{RGB}{176,190,197} 
\definecolor{backgroundSeventh}{RGB}{255,245,157} 

\setcopyright{none}
\copyrightyear{2020}
\acmYear{2020}
\acmDOI{XX.XXXX/XXXXXXX.XXXXXXX}

\acmConference[Preprint]{Preprint}{Date}{Virtual}
\acmBooktitle{Preprint, Virtual}

\acmPrice{XX.XX}
\acmISBN{XXX-X-XXXX-XXXX-X/XX/XX}

\begin{document}

 \title{\sol: {D}eriving {A}utomatically {V}erilog from {E}nglish}



 \author{Hammond Pearce}
 \affiliation{%
   \institution{New York University}
   \streetaddress{370 Jay Street}
   \city{Brooklyn}
   \country{USA}}
 \email{hammond.pearce@nyu.edu}

 \author{Benjamin Tan}
 \affiliation{%
   \institution{New York University}
   \streetaddress{370 Jay Street}
   \city{Brooklyn}
   \country{USA}}
 \email{benjamin.tan@nyu.edu}

 \author{Ramesh Karri}
 \affiliation{%
   \institution{New York University}
   \streetaddress{370 Jay Street}
   \city{Brooklyn}
   \country{USA}}
 \email{rkarri@nyu.edu}

\ignore{
\author{Author(s)}
\affiliation{%
  \institution{Institution}
  \streetaddress{Address}
  \city{City}
  \country{Country}}
\email{Email}

\author{Author(s)}
\affiliation{%
  \institution{Institution}
  \streetaddress{Address}
  \city{City}
  \country{Country}}
\email{Email}

\author{Author(s)}
\affiliation{%
  \institution{Institution}
  \streetaddress{Address}
  \city{City}
  \country{Country}}
\email{Email}
}
\begin{abstract}

While specifications for digital systems are provided in natural language, engineers undertake significant efforts to translate them into the programming languages understood by compilers for digital systems. Automating this process allows designers to work with the language in which they are most comfortable ---the original natural language--- and focus instead on other downstream design challenges. 
We explore the use of state-of-the-art machine learning (ML) to automatically derive Verilog snippets from English via fine-tuning GPT-2, a natural language ML system.
We describe our approach for producing a suitable dataset of novice-level digital design tasks and provide a detailed exploration of GPT-2, finding encouraging translation performance across our task sets ($94.8$\,\% correct), with the ability to handle both simple and abstract design tasks. 



\end{abstract}

\maketitle

\section{Introduction}
\label{sec:intro}

\begin{figure}[b]
\begin{tcolorbox}[fontlower=\footnotesize, fontupper=\footnotesize,top=3pt,bottom=2pt,left=3pt,right=3pt,]
{TASK: Write sequential code for a call button (e.g., in an airplane or hospital). If the call button 'b' is pressed (= 1) then the call light 'l' should turn on (= 1).
The output call light 'l' should turn off (= 0) when the synchronous cancel button 'r' is pressed (= 1). }\\ \vspace{-4mm}
\tcblower  \vspace{-2mm}
{RESULT:}
\begin{lstlisting}
// assume clock clk
reg l;
always @(posedge clk) begin
 if(r) begin
  l <= 0;
 end else if(b) begin 
  l <= 1;
 end
end
\end{lstlisting}
\vspace{-7mm}
\end{tcolorbox}
\vspace{-4mm}
\caption{Example "Task" and Corresponding Verilog\label{fig:example}}
\vspace{-10mm}
\end{figure}

In pursuit of simplifying and acceleration digital design, a machine-driven design flow with ``no humans in the loop'' is a long-term goal of projects such as OpenROAD\footnote{https://theopenroadproject.org/}.
Typically, the starting point is human-prepared hardware specifications in a \ac{HDL} such as Verilog.
However, manually producing \ac{HDL} to match a given specification (e.g. in \autoref{fig:example}) requires significant domain knowledge and is challenging to write error-free. 
As such, there is an opportunity for automatic translation to increase productivity and reduce the burdens on human designers. 
Given successful adoption of \ac{ML} throughout the \ac{IC} \ac{CAD} flow (e.g,~\cite{servadei_accurate_2019,yu_developing_2018,kahng_machine_2018}), we are motivated to investigate if state-of-the-art \ac{ML} can help in even earlier design stages. 

\ac{ML} has recently made great strides in \ac{NLP}. 
Advances in \ac{DL} have included new architectures such as LSTMs~\cite{sundermeyer2012lstm}, RNNs~\cite{liu2016recurrent}, and Transformers~\cite{vaswani_attention_2017}. 
These architectures have led to models such as BERT~\cite{devlin2018bert} and GPT-2~\cite{radford2019language} which demonstrate capability in language modelling, language translation (e.g., English to French), reading comprehension/understanding (e.g., answering questions from the CoQA~\cite{reddy2019coqa} dataset), and information storage/retrieval. 
In fact, GPT-2 made headlines~\cite{hern2019new} for initially being "too dangerous" to release given the "quality" of its text generation. 
\emph{Can we harness this power to produce hardware from task descriptions (like in~\autoref{fig:example})?} 

Towards the goal of fully automated design from natural language, we investigate the adaptation of a pre-trained natural language model to perform English to Verilog "translation". 
Using transfer learning~\cite{pan_survey_2010}, we fine-tune the recently presented GPT-2 for this task by training it on a custom dataset of Task/Result pairs, as in~\autoref{fig:example}. 
The tasks are somewhat akin to novice-level "textbook" problems (i.e., similar to those found in a classic textbook~\cite{vahid_digital_2010}).
We validate our approach by presenting a set of "unseen" tasks to translate and measure the quality of output. 
Our contributions are:
\begin{itemize}[noitemsep,topsep=0pt,leftmargin=*]
\item \sol, a pre-trained GPT-2 model that can translate natural language into Verilog implementation.
\item A method to automatically generate a large quantity of English specification, Verilog pairs for fine-tuning \sol.
\item Exploration and evaluation of fine-tuning \sol.
\item Rating ~\sol~ in translating complex \emph{descriptive} tasks besides those presented in simple prescriptive forms.
\end{itemize}
The rest of the paper is as follows. 
\autoref{sec:related} provides background and discuss related work. 
\autoref{sec:experiment} describes our experimental approach and dataset preparation. 
\autoref{sec:results} presents the results of our experimental investigation. 
\autoref{sec:conclusions} concludes.

\section{Background and Related Work}
\label{sec:related}

\textbf{ML-CAD.}
\ac{ML} techniques, including \ac{DL} have shown promising results across numerous applications, including across the \ac{CAD} domain. 
Recent work spans the design flow, from early-stage hardware cost estimations~\cite{servadei_accurate_2019}, through logic synthesis~\cite{yu_developing_2018}, and physical design~\cite{kahng_machine_2018}. 
We explore the use of transfer learning~\cite{pan_survey_2010} to teach a \ac{DL}-based model to produce Verilog by framing it as a machine translation problem. 
Transfer learning provides the ability to learn new tasks without large quantities of labelled data in a target domain. 

\textbf{GPT-2.}
We use GPT-2~\cite{radford2019language} as our starting point, given its state-of-the-art performance in zero-shot task settings. 
GPT-2 is based on the decoder part of the Transformer, a neural network encoder-decoder architecture with a self-attention mechanism~\cite{vaswani_attention_2017}.  
At the core of the GPT-2 approach is language modelling, which can be framed as an unsupervised distribution estimation from some set of examples $(x_1, x_2, ..., x_n)$, where each example is composed of variable length sequences of symbols $(s_1, s_2, ..., s_n)$~\cite{radford2019language}. 
This statistical model of language is thus the joint probability distribution of the symbols in the language (as the product of the conditional probabilities for each symbol given the preceding sequence~\cite{bengio_neural_2003}). 
Put simply, the model learns to answer the following: \emph{given some sequence of symbols, what is the most likely next symbol in the sequence?} 

Different tasks can be specified in a language itself, e.g., \emph{\{"translate to french", "english text", "french text"\}}~\cite{radford2019language}.
Radford \textit{et al.} speculate that a model with sufficiently large capacity can learn to perform tasks demonstrated in natural language without explicit supervision. 
In other words, given a general system which produces $p(output|input)$, a condition can be introduced to model some task $p(output|input,task)$. 
By training GPT-2 on a large, unlabelled dataset ($\sim$8 million webpages), Radford \textit{et al.} demonstrated the the trained model could perform well on numerous tasks without fine-tuning. 
The trained model then provides a good starting point for performance in specific tasks following fine-tuning~\cite{radford_improving_nodate}. 
Fundamentally, GPT-2's pre-trained, implicit capability to process natural language can be directed towards specific tasks. 
We attempt to harness this capability by fine-tuning GPT-2 for translating natural language descriptions to Verilog. 





\textbf{Natural Language $\rightarrow$ Code.}
The challenges in translating specifications into computer code has driven research in natural language programming~\cite{mihalcea2006nlp}. 
%
Recent work has shown that there is a finite limit to the number of unique ways one can express certain programming structures (e.g. \emph{for}-loops) in natural language, and as such it is possible to extract this information and transform it into its corresponding computer code \cite{mihalcea2006nlp}. 
Other related works use \ac{NLP} techniques, including rule-based processing, for formal system modeling~\cite{drechsler2012generating},
generating hardware assertions~\cite{harris2016glast}, and for enhancing documentation by automatically extracting software development tasks and associating them with the relevant paragraphs~\cite{treude_extracting_2015}. 
While showing promising results, there are limitations on how flexible the natural language descriptions can be with respect to structure. 
Earlier work involves designing separate components to perform specific tasks such as identifying "steps", "loops", and "comments" from natural text~\cite{mihalcea2006nlp}. 
To our knowledge, \ac{DL} techniques to generate \ac{HDL} from natural language have not been explored. 









\section{Fine-tuning GPT-2 for Verilog}
\label{sec:experiment}

\subsection{Problem definition}

In this work, we focus on an early-stage \ac{CAD} problem: interpreting a high-level, informal description of functionality and producing the corresponding concrete specification. 
For small designs, designers can craft an RTL specification directly after identifying the necessary inputs, outputs, and the relationships between them from a short description of a task.
While previous works use algorithmic approaches such as parse-tree generation and sub-tree matching~\cite{zhao_automatic_2019} to identify the salient elements of the natural language description for populating templates, we re-cast the problem holistically as \textit{translation}.
As we describe next, we prepare examples of task descriptions with varying descriptiveness, and examine GPT-2's ability to produce Verilog after transfer learning~\cite{pan_survey_2010}. 
\begin{figure}[t]
    \centering
    \includegraphics[width=\columnwidth]{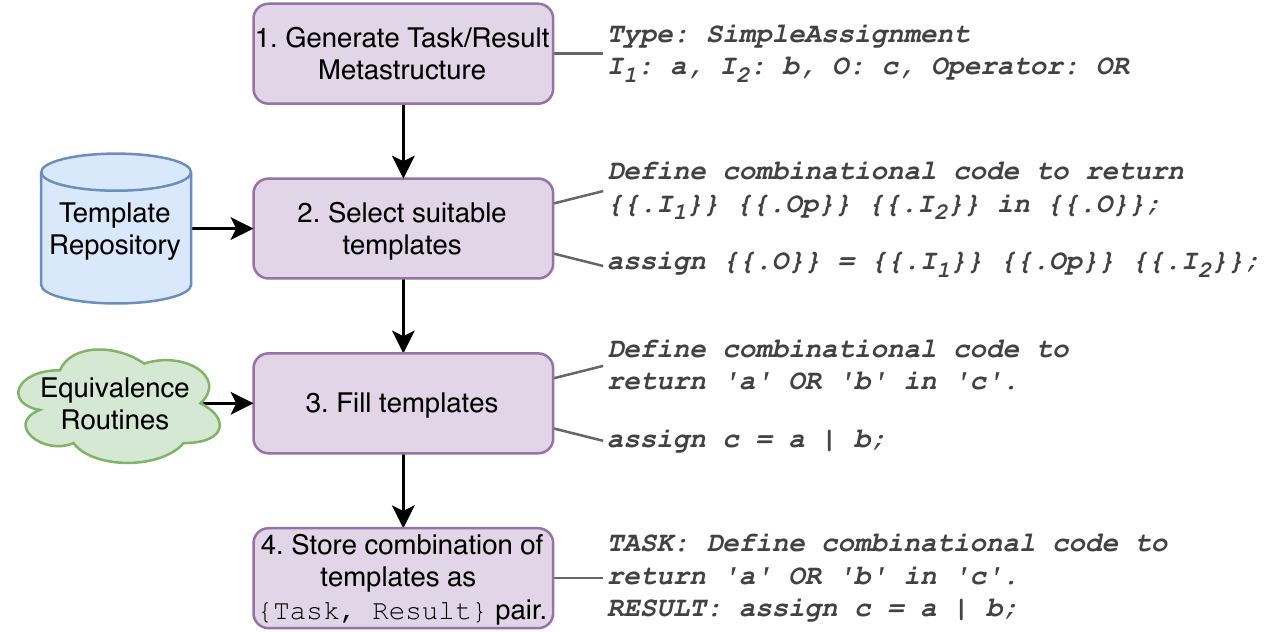}
    \caption{The Task/Result Generation Process \label{fig:task-result-gen-process}}
    \vspace{-5mm}
\end{figure}

\subsection{Dataset Preparation}
\label{sec:dataset-prep}

\begin{table*}[tb]
\caption{Template-based Dataset Information. (pX $\rightarrow$ prescriptive; dX $\rightarrow$ descriptive; X is the task type) \label{tbl:dataset}}
\vspace{-4mm}
\small
\renewcommand{\arraystretch}{0.2}
\begin{tabular}{@{}C{0.2cm}C{0.5cm}C{1cm}C{1cm}C{1.5cm}L{6cm}p{5.5cm}@{}}
\toprule
\multicolumn{2}{c}{Task}  & \# for Training & \# Non-Training & Samples / Template & Example of Task in English & Model Verilog \vspace{-1mm}\\ \midrule
\multirow{2}{*}{\rotatebox[origin=c]{90}{Assignment (a)\hspace{-2mm}}} & pa & 17 & 2 & 2000 & {\it Given inputs `a' and `b', take the nor of these and return the result in `c'.} & {\begin{lstlisting}[aboveskip=0pt,belowskip=-12pt,frame=none]
assign c = !(a | b);
\end{lstlisting}}\\ \cmidrule(r){2-7}
& da & 3 & 1 & 4000 & {\it A house has three active-low alarm detector triggered sensors `a', `b', `c'. Write combinatorial logic for a active-high light `l' which activates when any of the detectors are triggered.}  & {\begin{lstlisting}[aboveskip=-5pt,belowskip=-12pt,frame=none]
assign l = !(a & b & c);
\end{lstlisting}} \\ \midrule
\multirow{2}{*}{\rotatebox[origin=c]{90}{Register (r)}} & pr & 9 & 2 & 3000 & {\it Define a 4-bit register `q' with input `a' nand `b', enable `e' defined as `b' xnor `r', an asynchronous reset `r', and a clock `c'.} &  
{\begin{lstlisting}[aboveskip=-15pt,belowskip=-25pt,frame=none]
assign e = b ^ r; reg q;
always @(posedge c or posedge r) begin
if(r) begin  q <= 0; end 
else if(e) begin  q <=!(a & b); end
end;
\end{lstlisting}}
\\ \cmidrule(r){2-7}
& dr & 3 & 1 & 4000 & {\it Design the code for an alarm system. When the panic mode `m' is selected (= 0) the alarm system `a' should activate (= 1) and should only deactivate (= 0) when the active-low synchronous cancel button `c' is selected (= 1).} &  
{\begin{lstlisting}[aboveskip=-20pt,belowskip=-25pt,frame=none]
//assume clock clk
reg a;
always @(posedge clk) begin
if(c) begin  a <= 0; end 
else if(!m) begin  a <= 1; end
end
\end{lstlisting}}
\\ \midrule
\rotatebox[origin=c]{90}{Sequence Generator (g)\hspace{-1mm}} & pg & 4 & 2 & 4000 & {\it Define sequential code which will produce the repeating sequence [0, 1, 0] on output `u'. It should advance on clock `c' whenever enable `e' is present, and a synchronous reset `r' should reset the sequence back to the first element.} &  
{\begin{lstlisting}[aboveskip=-25pt,belowskip=-20pt,frame=none]
enum {s0, s1, s2} state; reg u;
always @(posedge c) begin
if(s) begin state <= s0; u <= b0; end 
else begin 
unique case (state) 
s0: if(e) begin state <= s1; u <= b0; end 
s1: if(e) begin state <= s2; u <= b1; end 
s2: if(e) begin state <= s0; u <= b0; end 
endcase 
end
\end{lstlisting}}
\\ \midrule
\rotatebox[origin=c]{90}{Multi-task (M-T)\hspace{-6mm}}& -- & N/A & N/A  & 5250 & {\it Write a 6-bit register `ar' with input defined as `gv' modulo `lj', enable `q', synchronous reset `r' defined as `yxo' greater than or equal to `m', and clock `p'. A vault door has three active-low secret switch pressed sensors `et', `lz', `l'. Write combinatorial logic for a active-high lock `s' which opens when all of the switches are pressed. Write a 6-bit register `w' with input  `se' and `md', enable `mmx', synchronous reset `nc' defined as `tfs' greater than `w', and clock `xx'.} & 
{\begin{lstlisting}[aboveskip=-45pt,belowskip=-20pt,frame=none]
assign r = yxo >= m; reg [5:0] ar;
always @(posedge p) begin
 if(r) begin ar <= 0; end 
 else if(q) begin ar <= gv % lj; end
end
assign s = !(et | lz | l); 
assign nc = tfs > w; reg [5:0] w;
always @(posedge xx) begin
 if(nc) begin w <= 0; end 
 else if(mmx) begin w <= se & md; end
end
\end{lstlisting}}
\\ \bottomrule
\end{tabular}%
\vspace{-3mm}
\end{table*}



In this work, we fine-tune GPT-2 to produce \sol, aiming for the ability to translate natural language (i.e., English) into Verilog. 
GPT-2 is designed to process contiguous text sequences, so we adopt the approach proposed in~\cite{radford_improving_nodate}, to represent the English--Verilog translation task as an ordered sequence 
in the format \texttt{`TASK: <English Text> RESULT: <Verilog Code>'}.

Open-source Verilog code can be found online, but is unstructured, with varying quality and complexity. 
For this initial study, we design a custom dataset generation tool inspired by the sort of template-based, random auto-marking \emph{Q\&A} systems used in teaching settings (e.g., the OASIS Question Engine\footnote{https://www.oasisqe.com/}).
Rather than produce thousands of Task/Result pairs manually, we prepare several natural language \emph{templates} which encapsulate different task scenarios. 
An example generation process is shown in \autoref{fig:task-result-gen-process}.

In step (1) our tool generates a Task/Result \emph{metastructure}, a descriptor for the type of task (e.g., an assignment) and relevant information for that task (e.g., variable names, operators).
Possible metastructure tasks include combinational signal \emph{assignments}, \emph{registers}, \emph{sequence generators}, or a multi-set of these. 
Then, in step (2), the tool randomly chooses a suitable template for the task that encapsulates all information in English and Verilog.
In step (3), the tool ``fills in'' these templates, translating arguments where necessary (e.g. \texttt{OR} operator is `\emph{or}' in English and `\texttt{|}' in Verilog). Finally, in step (4), the tool saves the generated Task/Result pair. 

Structurally, we organise our templates into the different task classes they describe---(combinational) assignments, registers, and sequence generators.
We then categorise them further as either \emph{prescriptive} or \emph{descriptive}. 
\textbf{Prescriptive templates} are like the example presented in \autoref{fig:task-result-gen-process}. 
We conjecture that these should be trivial to translate---simple substitutions and word-reordering is all that is required to convert from the English to Verilog.
\textbf{Descriptive templates}, meanwhile, are more like the example presented in \autoref{fig:example}. 
They are more complex to translate, and a human designer would implicitly perform intermediate steps---such as understanding that a given input is being used as an enable signal or as a reset.
\textbf{Multi-task} templates are random concatenations of two to four assignment/register templates. 
\autoref{tbl:dataset} provides additional examples of the different task types generated from the various templates. 

While at first glance this template-based approach for dataset generation might appear to restrict \sol's ability to generalize over English descriptions, this dataset is only used for \emph{fine-tuning} the language model. As GPT-2 is pre-trained over the large WebText dataset~\cite{radford2019language}, we theorize that \sol~ should retain at some ability to process natural language features such as  synonyms and different word/clause orders. To validate this hypothesis, we hold-out a subset of templates for use during testing and evaluation. 
\autoref{tbl:dataset} has information about the final dataset, including the number of "Trained" and "Non-Trained" (held-out) templates for all task types. 


In our evaluation, we initially query \sol~ with new task instances based on Trained templates to observe its baseline ability to perform "familiar" tasks (i.e., produce Verilog from English descriptions that are similar to the training data). 
To study generalizability of the approach, we query \sol~ with new task instances based on Non-Trained templates, i.e., such Task/Result pairs 
are presented to the language model during validation.

While the number of templates might appear low in certain cases (e.g., \# of Descriptive vs. Prescriptive assignments), the task instances of the given templates vary significantly from each other due to the addition or omission of optional clauses in the natural text during data generation.  A template that describes a register design task may have a clause describing a reset signal, and if the template is used for a metastructure with no reset signal, that entire clause is omitted.
As such a given template identifier refers only to the overall sentence structure used in a Task, the unique pattern of compulsory words within that template, such as introductory remarks (e.g. ``Describe combinatorial logic to...''), and individual words used within that template (e.g. conjunctions, prepositions).
Descriptive templates have randomly generated \emph{settings} such as ``an attendant call button''. These are generated from the cascaded sub-templates, increasing the entropy of each individual Task/Result pair.
Register and Sequence Generator templates are allowed to recursively define the basic template (prescriptive assignments).
A register might define a signal (e.g. an enable) as a function (e.g. `a' nand `b') rather than as a pre-set input (e.g. `c').

Multi-tasks combine other types of tasks and are difficult to categorise. We randomly generate 5,250 multi-task samples, of which 5000 are used for fine-tuning. We discuss details in \autoref{sec:results-multi-tasks}.

\subsection{Experimental Platform}
After we generate a suitable corpus of Task/Result pairs according to the method described in \autoref{sec:dataset-prep}, we fine-tune the 345 million parameter GPT-2 model on a high-performance computing node with 2 Intel Xeon E5-2698 v4 @ 2.20GHz cores, 20~GB of RAM, and an NVIDIA V100 32~GB graphics card over all categories of Task/Result pairs simultaneously (i.e. the same trained model is used to decode each type of Task). Our fine-tuning script is modified from \cite{aitextgen}. 
We use the Python programming environment, with \emph{pytorch} version 1.5.0, \emph{tensorflow} version 2.2, and \emph{aitextgen} version 0.2.3. Underlying these we use \emph{cuda} and \emph{cudnn} version 10.1.

To fine-tune GPT-2, we leave the hyper-parameters at their suggested defaults (\emph{learning rate} 1e-4, \emph{weight decay} 0.05, \emph{adam epsilon} 1e-8) and perform fine-tuning for 7500 steps.
The training data covers a random sample of 95\% of the generated samples of each Trained template category, with 5\% held back for evaluating the model.
To evaluate model "goodness", we use the same computing resources as for training and use default GPT-2 output generation parameters (\emph{temperature} 0.7, \emph{top\_p} 0.9, and \emph{top\_k} 0/disabled).

\section{Experimental Investigation}
\label{sec:results}

\subsection{Overview}
\label{sec:res-overview}
The purpose of this work is to explore the potential for general-purpose language models in translating system specifications provided in English to their hardware implementations in the Verilog \ac{HDL}.
As such we are interested in measuring the quality of the generated Verilog.
This raises an obvious question---how should one define "quality"? 
In this work we are interested in a language model which can perform design tasks of a similar difficulty to those posed in a textbook~\cite{vahid_digital_2010}.

However, there are no automated systems to quantify how well a specification has been implemented in its corresponding Verilog if it is "almost" correct. 
Formal equivalence check is an option, but requires that the design is at least syntactically compliant. 
This presents a challenge as we wish to quantify the quality of \sol's Verilog generation. 
However, given that we generate Task/Result pairs with a template engine, we have a baseline `canonical' response that we can compare \sol's output against.
This allows us to introduce the equivalence between the two generators as a measure of quality, discussed in \autoref{sec:equality}. 
Where \sol's output is not equivalent, we manually examine the result qualitatively. 

An important part of our evaluation is to examine \sol's performance over unfamiliar texts.
Otherwise, it could be argued that the language model has simply learned a kind of pattern recognition over the Task/Result pairs, and is just using string relocation techniques to score highly during validation.
If this notion were applied to a student, we might say that they had learned to produce Verilog by rote, rather than through understanding. 

This examination is provided through the Non-Trained Templates. 
Recall that these are unfamiliar to \sol, i.e., they were not seen during fine-tuning,
and \sol~ has had no opportunity to learn/memorize their syntax and structure.
%
We seek insight from \sol's performance over these tasks as evidence that the GPT-2 language model offers promise for our intended translation purpose.
%




\subsubsection{A measure of equality.} 
\label{sec:equality}
There are numerous ways to implement a given specification in any programming language. 
Take the example from \autoref{fig:task-result-gen-process}: while it provides the correct answer as \texttt{assign c = a~|~b;}, it could be equivalently specified as \texttt{assign c = b | a;}.
This becomes even more of an issue when implementing larger and more complex and descriptive specifications.

While there are ways of quantifying identical code (e.g., comparing abstract syntax trees), we opt, for a simpler comparison of \sol's outputs against the template tool using a sequence equivalence metric. This is because the generated Verilog code should be relatively short and simple.
More precisely, we define \textit{correctness} of the generated text as its distance to the template-provided ``correct'' answer (excluding white-space characters from both) as measured by their Ratcliff-Obershelp similarity~\cite{ratcliff1998ratcliff}.
This means that if \sol~ returns \texttt{assign c = a | b;} as the correct answer to the prompt in \autoref{fig:task-result-gen-process}, it scores $1.00$---i.e., the result is fully correct.
However, despite being functionally equivalent, a result of \texttt{assign c = b | a;} scores only $0.833$.

While this metric is simple, 
manual inspection of the results that did not have the expected score of $100$, revealed no examples where \sol~ had performed small but functionally equivalent changes (e.g., inverting the order of variables compared to their order in the specification). That the output has a deterministic ordering to the variables is not a surprising result, as the template engine that \sol~ is fine-tuned from has a deterministic order to the Verilog code that it produces. We provide insights from our investigation in three parts: \sol's performance on prescriptive (\autoref{sec:translation-prescriptive}), descriptive (\autoref{sec:translation-descriptive}), and multi tasks (\autoref{sec:results-multi-tasks}).

\subsection{Translation of Prescriptive Specifications}
\label{sec:translation-prescriptive}

\begin{table}[t]
\centering
\caption{Testing \sol~ on Prescriptive Tasks\label{tbl:res-prescriptive}}
\vspace{-4mm}
\resizebox{\columnwidth}{!}{%
\begin{tabular}{C{0.7cm}C{2cm}C{1.5cm}C{1.6cm}C{1.5cm}C{1.5cm}}
\hline
\textbf{Type}       & \textbf{Template Name} & \textbf{\# Trained} & \textbf{\# Validated} & \textbf{\# Correct} & \textbf{Avg. Error R-O} \\ \hline
\multirow{19}{*}{\rotatebox{90}{Assignment}}
    & pa00   & 1900  & 100     & 99   & 0.947   \\ \cline{2-6} 
    & pa01   & 1900  & 100     & 100   & --   \\ \cline{2-6} 
    & pa02   & 1900  & 100     & 100   & --   \\ \cline{2-6} 
    & pa03   & 1900  & 100     & 100   & --   \\ \cline{2-6} 
    & pa04   & 1900  & 100     & 100   & --      \\ \cline{2-6} 
    & pa05   & 1900  & 100     & 100   & --   \\ \cline{2-6} 
    & pa06   & 1900  & 100     & 97   & 0.951      \\ \cline{2-6} 
    & pa07   & 1900  & 100     & 100   & --   \\ \cline{2-6} 
    & pa08   & 1900  & 100     & 100   & --   \\ \cline{2-6} 
    & pa09   & 1900  & 100     & 100   & --      \\ \cline{2-6} 
    & pa10   & 1900  & 100     & 100   & --   \\ \cline{2-6} 
    & pa11   & 1900  & 100     & 100   & --   \\ \cline{2-6} 
    & pa12   & 1900  & 100     & 100   & --   \\ \cline{2-6} 
    & pa13   & 1900  & 100     & 100   & --      \\ \cline{2-6} 
    & pa14   & 1900  & 100     & 99   & 0.947       \\ \cline{2-6} 
    & pa15   & 1900  & 100     & 100   & --   \\ \cline{2-6} 
    & pa16   & 1900  & 100     & 100   & --   \\ \cline{2-6} 
    & \textbf{pa17}   & \textbf{0}    & \textbf{100}     & \textbf{95}   & \textbf{0.956}      \\ \cline{2-6} 
    & \textbf{pa18}   & \textbf{0}    & \textbf{100}     & \textbf{98}   & \textbf{0.898}      \\ \hline
\multirow{15}{*}{\rotatebox{90}{Register}}  
    & pr00   & 2850 & 150    & 148  & 0.981   \\ \cline{2-6} 
    & pr01   & 2850 & 150    & 149  & 0.993   \\ \cline{2-6} 
    & pr02   & 2850 & 150    & 149   & 0.973      \\ \cline{2-6} 
    & pr03   & 2850 & 150    & 150   & --      \\ \cline{2-6} 
    & pr04   & 2850 & 150    & 148   & 0.990      \\ \cline{2-6} 
    & pr05   & 2850 & 150    & 147  & 0.982   \\ \cline{2-6} 
    & pr06   & 2850 & 100    & 148   & 0.993     \\ \cline{2-6}     
    & pr07   & 2850 & 150    & 149   & 0.983      \\ \cline{2-6} 
    & pr08   & 2850 & 150    & 150   & --      \\ \cline{2-6} 
    & pr09   & 2850 & 150    & 150   & --      \\ \cline{2-6} 
    & \textbf{pr10}   & \textbf{0}    & \textbf{150}     & \textbf{149}   & \textbf{0.960}   \\ \cline{2-6} 
    & \textbf{pr11}   & \textbf{0}    & \textbf{150}     & \textbf{147}   & \textbf{0.965}      \\ \hline
\multirow{6}{*}{\rotatebox{90}{Seq. Generator}} 
    & pg01   & 3800 & 200    & 200  & --      \\ \cline{2-6} 
    & pg02   & 3800 & 200    & 199  & 0.996      \\ \cline{2-6} 
    & pg03   & 3800 & 200    & 200  & --      \\ \cline{2-6} 
    & pg04   & 3800 & 200    & 197  & 0.984      \\ \cline{2-6} 
    & \textbf{pg05}   & \textbf{0}    & \textbf{200}     & \textbf{200}   & \textbf{--}      \\ \cline{2-6} 
    & \textbf{pg06}   & \textbf{0}    & \textbf{200}     & \textbf{143}   & \textbf{0.889}      \\ \hline
\end{tabular}
}
\vspace{-5mm}
\end{table}

\sol's performance on prescriptive tasks is presented in \autoref{tbl:res-prescriptive}, with Non-Trained templates highlighted in \textbf{bold}. Each row contains information on the number of template samples used for fine-tuning, the number of template samples used for validation, the number \sol~ returned correctly, and (where applicable) the average Ratcliff-Obershelp (R-O) similarity of returned incorrect answers compared to the correct answer.

With regards to assignments, \sol~ performs  well on tasks based on Trained (e.g., \textit{pa00}\footnote{\textit{pa00} example: ``Put the result of `a' nand `b' in `c'.''}) templates, getting 99.7\,\% of all samples correct across this validation category.
It performs slightly worse on tasks drawn from Non-Trained templates (e.g., \textit{pa18}\footnote{\textit{pa18:} ``Assign into output `c' the result of `a' xor `b'.''}), scoring 96.5\,\% correct.
\sol~ scores well on Trained register templates (e.g., \textit{pr00}\footnote{\textit{pr00:} ``Define a 8-bit register `a' with input `a' defined as `b' and `c', enable `e', and clock `c'.''}) (99.2\,\% correct).
Likewise \sol~ performed well with the Non-Trained Templates in this category (e.g. \textit{pr11}\footnote{\textit{pr11}: Given input 'a', enable 'e' defined as 'd' nxor 'f', an asynchronous reset 'r' (being 'x' or 'y') make a 7-bit register 'q'.}), with 98.7\,\% correct.
While \sol~ did well in Trained Sequence Generators (e.g. \textit{pg01}\footnote{\textit{pg01:} ``Define sequential code which will produce the repeating sequence [00, 10, 10] on the 2-bit output `q'. It should advance on each tick of a clock `c' whenever enable defined as `a' nxor `b' is present.''}) with 99.5\,\% correct across the samples, it performed poorly with the Non-Trained template \textit{pg06}\footnote{\textit{pg06:} ``Produce a design that generates a 3-bit output `uy' with the sequence: [110, 100, 101, 100]. The output changes with each rising edge of a clock if the enable signal `a' less than `b' is asserted. Whenever an asynchronous reset `r' is asserted, the design should output the first element of the sequence.''}, bringing the overall percentage correct for Non-Trained Templates down to 85.6\,\%.
 
\textbf{Discussion.} 
One would expect \sol~ to perform well on tasks produced from Trained templates, given that 
these most resemble the training data. 
This held true for all three major categories.
One might also expect that \sol~ would perform worse on task prompts generated from Non-Trained templates in comparison to prompts generated from the Trained templates. 
Our hypothesis is that the GPT-2 pre-training should allow \sol~ to generalise and produce the correct Verilog even in unseen tasks. 

This holds for Assignments and Registers, but did not entirely hold with the Non-Trained Sequence Generator templates, specifically with \textit{pg06}.
Closer investigation of this template revealed that almost all of \sol's errors (>95\,\%) stem from mis-classification of enable and reset signals. This was unexpected as \sol~ did not have this issue over tasks based on any other Sequence Generator template.
One theory is that the issue may stem from the difference between \textit{pg06} and the other templates---perhaps it is \emph{too} unique. 
To evaluate this, we compared the the R-O similarity of templates \textit{pg05} (which scored 100\,\%) and \textit{pg06} with the Trained \textit{pg} templates.
We found that \textit{pg05} was closest to \textit{pg01} (similarity $0.820$), whereas \textit{pg06} was closest to \textit{pg03} (similarity $0.777$).
These numbers are similar enough that we would have expected \textit{pg06} to score better. Further formal analysis is an avenue for our future work. It is likely that providing a greater variety of Sequence Generator templates during training would help \sol~ produce more accurate results.




\begin{table}[tb] 
\centering
\caption{Testing DAVE on Descriptive and Multi- Tasks\label{tbl:res-descriptive}}
\vspace{-4mm}
\resizebox{\columnwidth}{!}{%
\begin{tabular}{C{0.7cm}C{2cm}C{1.5cm}C{1.6cm}C{1.5cm}C{1.5cm}}
\hline
\textbf{Type}                       & \textbf{Template Name} & \textbf{\# Trained} & \textbf{\# Validated} & \textbf{\# Correct} & \textbf{Avg. Error R-O} \\ \hline
\multirow{5}{*}{\rotatebox{90}{Assign.}}
& da00                   
& 3800                                          
& 200                                             
& 200                                           
& --                           
\\ \cline{2-6} 
& da01                   
& 3800                                          
& 200                                             
& 199                                           
& 0.952                           
\\ \cline{2-6} 
& da02                   
& 3800                                          
& 200                                             
& 196                                           
& 0.956                           
\\ \cline{2-6} 
& \textbf{da03}                   
& \textbf{0}                                          
& \textbf{200}                                             
& \textbf{200}                                           
& \textbf{--}                           
\\ \hline
\multirow{5}{*}{\rotatebox{90}{Register}}
& dr00 
& 3800
& 200                                            
& 200                                           
& --                                             
\\ \cline{2-6} 
& dr01                   
& 3800                                         
& 200                                            
& 195                                          
& 0.985                           
\\ \cline{2-6} 
& dr02                   
& 3800                                         
& 200                                            
& 199                                          
& 0.992                           
\\ \cline{2-6} 
& dr03                   
& 3800                                         
& 200                                            
& 198                                           
& 0.988    
\\ \cline{2-6} 
& \textbf{dr04}                   
& \textbf{0}                                         
& \textbf{200}                                            
& \textbf{196}                                           
& \textbf{0.987}                                      
\\ \hline 
\multirow{2}{*}{\rotatebox{90}{M-T}}
& Trained 
& 5000
& 250                                            
& 130                                           
& 0.907                                            
\\ \cline{2-6} 
& \textbf{Non-Trained}                   
& \textbf{0}                                         
& \textbf{250}                                            
& \textbf{103}                                           
& \textbf{0.817}                                      
\\ \hline
\end{tabular}
}
\vspace{-5mm}
\end{table}
\subsection{Translation of Descriptive Specifications}
\label{sec:translation-descriptive}

\autoref{tbl:res-descriptive} presents \sol's performance over Descriptive Tasks. 
While this category has fewer templates, each template has more opportunities for entropy due to the presence of optional clauses and implicit intermediate signals. 
We also design these templates to be more ``difficult''---they invoke requirements such as `active-high' and `active-low' qualifiers to their variables, terms that \sol~ needs to recognise and accommodate in the generated Verilog.

Somewhat surprisingly, \sol~ performs better on Descriptive Tasks than on the Prescriptive Tasks, with 99.2\,\% correct Assignments and 99.0\,\% Registers over the Trained Templates.
For the Non-Trained templates, the Assignments scored 100\,\% correct and Registers scored 98\,\%.
To check that this high score was not due to the Non-Trained templates \textit{da03} and \textit{dr04} being structurally similar to the Trained templates, we compare R-O similarities. \textit{da03} is most similar to \textit{da01}, with a score of $0.686$. \textit{dr04} is most similar to \textit{dr02}, with a score of $0.703$. 
While these values might seem high, consider the  Sequence Generator template \textit{pg06}, which scored $0.777$ yet \sol~ gave the correct answer only 71.5\,\% of the time.

\textbf{Discussion.} 
On a number of occasions, we were particularly impressed that \sol~ was able to derive the Boolean combinations for certain operations.
Take this example from \textit{da00}: ``A car has four active-low door open sensors `a', `b', `c', `d'. Write combinatorial logic for a active-low light `l' which illuminates when any of the doors are open.''
From that prompt, \sol~ is able to correctly generate the output \texttt{assign l = a \& b \& c \& d;}, i.e., it appears to associate `any' and `doors', as well as understand the relationship between `any' and the two `active-low' qualifiers.
Another example of \sol~ "understanding" keywords is the generated Verilog for \textit{dr00}, which we present in \autoref{fig:example}. 
\sol~ can correctly implement both synchronous and asynchronous resets, as well as infer clocks for memory elements when no clocks are explicitly specified.

\subsection{Translation of Multiple Tasks}
\label{sec:results-multi-tasks}

For insight into how \sol~ can handle the processing of multiple tasks simultaneously we also provided a multi-task metastructure consisting of 2-4 registers and assignments in a single Task prompt.
These are presented in \autoref{tbl:res-descriptive} under M-T.
We divide Multi-tasks into two broad categories---those made purely from Trained templates (of which 5000 were presented during the fine-tuning process), and those made only from Non-Trained templates.
Multi-tasks performed worse than the individual templates (Trained correct 52\,\% of the time, and Non-Trained 41.2\,\%).
Upon manual inspection, \sol~ was generating the correct Verilog structures and syntax in the outputs, usually only getting variable names/operators incorrect. This is reflected in the Average Error R-O, which is high given the answer lengths. It is likely that the difficulties \sol~ is facing with multi-tasks stem from the na\"{i}ve concatenation of tasks. In future we will explore multi-tasks where the "sub-tasks" are related. 


\subsection{Discussion and Limitations}

The results presented are promising.
\sol~ has shown clear ability to produce syntactically correct Verilog (in our tests, it rarely, if ever, produced outputs that could not compile---errors were almost always related to operator choice and/or variable names).
\sol~ is capable of producing code with complex relationships between inputs and outputs, and even with intermediate signals. In total, \sol~ returned the correct answer in \textbf{94.8\,\%} of all validation tests.

That said, our work has limitations.
Firstly, other than inferring clocks, we do not yet ask \sol~ to create a signal that was not already named or otherwise described (e.g., we never provide code such as "Output `a' nor `b'", it is always "Output `a' nor `b' \underline{in `c'}.").
Likewise, we never rely on any form of \emph{creativity} in the generated results---our training data suggests that there 
is only one path forward to the implementation for a given task template. 
That is, our templates had a \emph{many-to-one} relationship with the Verilog they described, despite there being different ways to express functionally identical Verilog. 
These are the focus of our ongoing  studies.

\sol~ inherits some \emph{technical} limitations of GPT-2: The model can only generate outputs of up to 1024 tokens (i.e., words, symbols). As longer snippets of code can potentially run into this limit, we had to limit certain inputs---sequence generators were capped at no more than 4 elements, and our multi-tasks were prevented from using long-winded descriptive register templates.

\section{Conclusions}
\label{sec:conclusions}

This paper set out to explore the potential use of \ac{ML} for translating natural language specifications into their corresponding Verilog \ac{HDL}.
We adopted the GPT-2 language model and fine-tuned it over a large number of English/Verilog Task/Result pairs to produce \sol. 
We investigated \sol's performance over sets of English to Verilog Tasks based on familiar and unfamiliar templates.
In general, \sol's performance exceeded our expectations and was able to produce Verilog in response to both simple, prescriptive prompts, as well show success in acquiring the advanced capabilities required to solve more descriptive settings.
Our future work will investigate the use of larger GPT-2 models for \sol, increasing the complexity and length of the tasks, and tuning \sol~ for specific tasks such as security assertion generation from natural language collateral. 

\section*{Try \sol~!}


\href{https://colab.research.google.com/drive/1aDSMDWL5hieB3_Th9ZdddDMAKQ2DjWxW}{\color{blue}{Click here for instructions to run \sol~ freely within Google Colab.}}




\bibliographystyle{ACM-Reference-Format}
\bibliography{IEEEabrv,references}

\end{document}